\documentclass[article,aps,pra,onecolumn,groupedaddress,showpacs]{revtex4}
\usepackage{amsmath}
\usepackage{amsfonts}
\usepackage{amssymb}
\usepackage{graphicx}%
\newtheorem{theorem}{Theorem}

\newtheorem{corollary}[theorem]{Corollary}

\newtheorem{lemma}[theorem]{Lemma}

\newenvironment{proof}[1][Proof]{\noindent\textbf{#1.} }{\ \rule{0.5em}{0.5em}}
\begin{document}

\title{Blind encoding into qudits}

\author{J. S. Shaari$^{a}$, M. R. B. Wahiddin$^{a,b}$, and S. Mancini$^{c}$}

\address{\smallskip $^{a}${\textit{Faculty of Science, International Islamic
University of Malaysia (IIUM), Jalan Gombak, 53100 Kuala Lumpur,
Malaysia}}
\\
$^{b}${\textit{Cyberspace Security Laboratory, MIMOS Berhad,
Technology Park Malaysia, 57000 Kuala Lumpur, Malaysia}}
\\
$^{c}${\textit{Dipartimento di Fisica, Universit\`a di
Camerino, 62032 Camerino, Italy}}}

\date{\today}

\begin{abstract}
We consider the problem of encoding classical information
into unknown qudit states belonging to
\emph{any} basis, of a maximal set of mutually unbiased bases, by one party and then decoding by another party who has perfect knowledge of the basis.
Working with qudits of prime dimensions,
we point out a no-go theorem that forbids `shift' operations on arbitrary unknown states.
We then provide the necessary conditions for reliable encoding/decoding.
\end{abstract}


\maketitle

\section{Introduction}

The idea of encoding and decoding classical information onto an unknown quantum state is
essentially related to transformations and measurements of vectors
state in a Hilbert space.

Suppose that Bob picks a qudit state from a given set
(a subset of a $d$-dimensional Hilbert space $\mathcal{H}_d$) and sends it to
Alice who is oblivious about the state. Alice is then expected to
encode classical information (one out of $d$ symbols)
by virtue of unitary transformations before
sending it back to Bob who should decode by gaining full information.
This task can be reliably accomplished once the initial set of states forms a basis of $\mathcal{H}_d$.
In fact, in such a case, Alice has simply to shift the incoming state into another of the basis
by an amount determined by the symbol she wants to encode,
while Bob has to measure in (project onto) the basis to retrieve Alice operation-symbol.

What happen if the initial set of states comprises more than one basis,
specifically a number of mutually unbiased bases (MUB) \cite{Ivan,Band}?

This problem is of fundamental interest in quantum cryptography.
For instance, blind encoding of classical information into states belonging to MUB is used
in two-way deterministic quantum key distribution
\cite{Cai,Deng,LM05,6dp,qutrit}. A crucial point to
note in these protocols is that security results from the ambiguity of bases
introduced by MUB. The endeavor for a
more secure protocol thus entails the problem of maximizing the
number of MUB over which encoding
operations perform equally and decoding may be done reliably.

The maximum cardinality of any set of MUB in $\mathcal{H}_d$ is exactly known  to be $d+1$ only
when $d$ is a prime power \cite{Ivan,Band}.
If, moreover, it is simply prime,
a straightforward construction of bases states exist \cite{Ivan,Band}.
We henceforth restrict our attention to qudits of prime dimensions.

Unfortunately, a nontrivial scenario already emerges
for qubits ($d=2$),
where the non existence of the universal-NOT  \cite{UNOT1,UNOT2}
forbids the ability to shift (flip)
arbitrary unknown qubit states while the unitary Pauli operators shift
qubits in only two out of three MUB.

This entails two main problems that we shall deal with in this paper:
\emph{i)} Can Alice blindly encode onto an unknown state of \emph{any} basis of a maximal set of MUB?
\emph{ii)} Can Bob efficiently decode the full information?  We discuss the first in
terms of a No-Go Theorem \cite{qutrit} which forbids the
shifting of one arbitrary pure state into another in Sec.II,
and we address the second problem in Sec.III by devising a specific protocol.
In order to quantify the figure of merit of this
protocol we consider, in Sec.IV, its efficiency within a communication framework
\cite{cabello1}. We then apply the protocol to some examples.
We reserve Sec.V for conclusions.


\section{The No-Go Theorem}

We start by considering a
 $d$-dimensional quantum system, i.e., a qudit.
 In its Hilbert space $\mathcal{H}_d$ we  choose a basis (computational basis)
 $\{\vert j\rangle\}$ labeled by elements $j\in\mathbb{Z}_d$.
Moving from the $d=2$ case (qubit), we can introduce generalized Pauli operators
$X$ and $Z$
such that
\begin{eqnarray}
&&X\left\vert j\right\rangle =\left\vert j+1\right\rangle,
\label{Xgen}\\
&&Z\left\vert j\right\rangle =\omega ^{j}\left\vert j\right\rangle,
 \label{Zgen}
\end{eqnarray}
with
\begin{equation}
\omega:=\exp \left( i\frac{2\pi }{d}\right).
\end{equation}
Generalized Pauli operators are unitary and satisfy the anticommutation
relation
\begin{equation}
ZX=\omega XZ.  \label{Anticomm}
\end{equation}

It is well known \cite{Ivan,Band} that in $\mathcal{H}_d$, with $d$ prime number, there are $d+1$ MUB and their states can be constructed as eigenstates of operators
\begin{equation}\label{bases}
XZ^0,\;XZ^1,\;XZ^2,\;\ldots,\;XZ^{d-1},\;Z.
\end{equation}
Now suppose that Bob picks one of these states, say $|\psi_t^k\rangle$ where $t=0,1,\ldots,d-1$ denotes the element within a basis and $k=0,1,\ldots,d$ denotes the basis.
Bob sends the state to Alice who is oblivious about it and she
wants to encode one of the symbols belonging to the alphabet $\mathcal{A}\equiv\{0,1,\ldots,d-1\}$.
She therefore requires a unitary shift operator $U$ such that
\begin{equation}
|\psi_t^k\rangle\stackrel{U^r}\longrightarrow|\psi_{t\oplus r}^k\rangle,
\end{equation}
where $\oplus$ stands for the sum mod $d$.

However, in considering a unitary operation that may shift qudit states in any MUB, we 
arrive at the following theorem (generalizing the one for qutrit \cite{qutrit} and extending the arguments for the nonexistence of Universal-NOT \cite{UNOT1,UNOT2}).

\begin{theorem}
\label{th1}
There is no unitary transformation that may shift between pure orthogonal states of any
 MUB of prime dimension.
\end{theorem}

\begin{proof}
We prove the theorem by \textit{reductio ad absurdum}. Let
us first consider the eigenvectors of operators $XZ^{k}$, $k=0,1...,d-1$
denoting $d$ different MUB, thus excluding the computational basis. They can be written as
\begin{equation}
\left\vert \psi_{t}^k\right\rangle=\frac{1}{\sqrt{d}}\sum_{j=0}^{d-1}\omega^{t\left( d-j\right) -ks_{j}}\left\vert j\right\rangle,
\end{equation}
where $s_{j}  =j+...+d-1$. We then assume the existence of a unitary transformation $U$  acting on the computational basis as
\begin{equation}
U\{\left\vert 0\right\rangle ,\left\vert 1\right\rangle
,...,\left\vert
d-1\right\rangle \}\rightarrow \{a_{0}\left\vert 1\right\rangle ,a_{1}
\left\vert 2\right\rangle ,...,a_{d-1}\left\vert 0\right\rangle\},
\label{Ucb}
\end{equation}
with $a_{i}\in
\mathbb{C}$ such that $|a_i|=1$ and $i=0,1...,d-1$. 

Without loss of generality, we may single out a state
$\left\vert \psi_{T}^k\right\rangle $ with index $T$ for any $k$ (basis) and
consider the operator $U$ acting on it
\begin{equation}
U\left\vert \psi_{T}^k\right\rangle= \frac{1}{\sqrt{d}}\sum
_{j=0}^{d-1}a_{j}\omega^{T\left( d-j\right)
-ks_{j}}\left\vert j+1\right\rangle.
\end{equation}
Then, for any $k$, the resulting vector should correspond to one of the other
vectors of the basis (orthogonal to the initial state) and 
we may write
\begin{eqnarray}
\label{coeff}
\frac{1}{\sqrt{d}}\sum_{j=0}^{d-1}a_{j}\omega^{T\left(  d-j\right)  -ks_{j}}\left\vert
j+1\right\rangle
=\frac{1}{\sqrt{d}}\sum_{j=0}^{d-1}\omega^{t\left(  d-j\right)-ks_{j}}\left\vert j\right\rangle,
\end{eqnarray}
with $t\neq T$ as necessary shift
requirement.  By equating the coefficients of the same states at both sides of Eq.(\ref{coeff}), we get
\begin{eqnarray}
a_{j} = \omega^{t\left(d-j-1\right)-T\left(d-j\right)+kj}.
\end{eqnarray}
Taking two indexes  $j$ and
$j'=j-i$ differing by an integer $i$, we have
\begin{align}
\frac{a_{j}}{a_{j'}}  =\omega^{(T+k-t)i},
\end{align}
and for $i=1$
\begin{align}
\frac{a_{j}}{a_{j+1}}=\omega^{-t+k+T},
\end{align}
which must be invariant with respect to $k$.
Since $t$ could be different for differing $k$s, we require
\begin{eqnarray}
\omega^{-t_{k}+k+T}=\omega^{-t_{k^{\prime}}+k^{\prime}+T},\quad\forall\, k\neq k'.
\end{eqnarray}
For every term $x\equiv k-t_{k},$ we may find a particular $k^{\prime}$
such that
\begin{align}
k^{\prime}=\left(  x+T\right)  \operatorname{mod}(d-1),
\end{align}
thus
\begin{align}
\omega^{x+T}  & =\omega^{-t_{k^{\prime}}+(x+T)\operatorname{mod}(d-1)+T},\\
1  & =\omega^{-t_{k^{\prime}}+T},\\
T  & =t_{k^{\prime}}.
\end{align}
The last equality contradicts the shift requirement $T\neq t_{k},\;\forall k$.
This completes the proof.
\end{proof}


\section{Requirements for Reliable Encoding/ Decoding}

Given the result of theorem \ref{th1}
together with the fact that a
unitary operator of the form $XZ^{l}$ may
shift eigenvectors of $XZ^{k}$ when $k\neq l$ (that is states in a number of $d$ MUB) \cite{Band}, 
we also have the following corollary.

\begin{corollary}
A unitary operation may shift a qudit state to an orthogonal one
in at most $d$ MUB.
\end{corollary}

This results in the impossibility for Alice to reliably encode on $d+1$ MUB.
An obvious example would see Bob sending a state which is the eigenvector of
$XZ^{k}$ and Alice cannot
encode a shift unless she knows the basis Bob used, that is $k$ (then she may use
$XZ^{l}$, $k\neq l$). Bob on the other hand
could not perfectly decode, since in the event he measured a shifted
state he could not discern between the $d$ different kinds of
transformations that would have resulted in the same evolution.
However, Bob may be able to lessen his uncertainty by sending more
qudits of differing bases and in the case that Bob may deduce
perfectly Alice's unitary transformation, the problems of
both encoding and decoding are solved. We
therefore propose the following lemma.

\begin{lemma}
A reliably blind encoding with $d+1$ MUB
needs to use strings of $d$ qudits 
from $d$ differing MUB.
\end{lemma}

\begin{proof}
Let us denote the unitary
transformation to shift a state in any basis except $k$th as
$U_{\overline{k}}$. The number of unitaries available
to shift  a qudit state (including identity) would be $d+2$. 
These are the unitaries $U_i$, $i=1,\ldots,d+1$, of Eq.(\ref{bases}) plus identity operation (say $U_{d+2}$). 
Bob's initial uncertainty about
unitary transformation $U$ amounts to
\begin{eqnarray}\label{HU}
H\left(  U\right)  =\log\left( d+2 \right).
\end{eqnarray}
where $H$ stands for the Shannon entropy.
Thus, the maximum information $I_{max}$ that Bob may gain is exactly given by Eq.(\ref{HU}).

Suppose that Bob prepares a qudit state and
that it undergoes Alice's unitary $U_i$. All $U_i$, $i=1,\ldots,d+2$, are equally probable, thus
\begin{eqnarray}
\Pr\left(U_i\right)=\frac{1}{d+2}.
\end{eqnarray}
Bob's subsequent measurement reveals whether the shift has taken place or not.
The latter happens when the unitary is either  $U_{\overline{k}}$ or $U_{d+2}$ (identity).
Thus, we have the following probabilities
\begin{eqnarray}
\Pr\left(  s=0\right)&=&\frac{2}{d+2},\\
\Pr\left(  s=1\right)&=&\frac{d}{d+2},
\end{eqnarray}
with $S$ is a binary random variable taking values to denote shift of the state ($s=0$ no shift, $s=1$ shift). Moreover,
\begin{equation}
\Pr\left(   U=U_{i}\mid s=0\right)=
\left\{\begin{array}{lcr}
\frac{1}{2}& &i=\overline{k},d+2\\
0& &i\neq\overline{k},d+2
\end{array}
\right. ,
\end{equation}
\begin{equation}
\Pr\left(  U=U_{i}\mid s=1 \right)=
\left\{\begin{array}{lcr}
0& &i=\overline{k},d+2\\
\frac{1}{d}& &i\neq\overline{k},d+2
\end{array}
\right. ,
\end{equation}

Then, Bob's uncertainty (about $U$) subsequent
to his measurement can be calculated
by using
\begin{eqnarray}
H\left(  U_{i}\mid S=s\right)  =-\sum\limits_{i}\Pr\left(  U_{i}\mid
S=s\right)  \log\left[  \Pr\left(  U_{i}\mid S=s\right)  \right],
\end{eqnarray}
so that Bob's a posteriori uncertainty results
\begin{eqnarray}\label{Hpost}
H_{post}\left(  U\right)   &&  =\sum_{s}\Pr\left( S=s\right)  H\left(  U_{i}\mid S=s\right)
\cr&& =\frac{d}{d+2}\left[  -\log\frac{1}{d}\right]
+\frac{2}{d+2}\left[ -\log\left(  \frac{1}{2}\right)  \right].
\end{eqnarray}
By using Eqs.(\ref{HU}) and (\ref{Hpost}) we get
\begin{eqnarray}
H\left(  U\right)  -H_{post}\left(  U\right)  <I_{max}.
\end{eqnarray}
If Bob had used two qudits of differing basis, $\overline{k}$ and $\overline{k'}$, with the encoding operation acting on both of them, 
we have to distinguish among four values of $s$ with probabilities
\begin{eqnarray}
\Pr\left(  s=00\right)&=&\frac{1}{d+2},\\
\Pr\left(  s=01\right)&=&\frac{1}{d+2},\\
\Pr\left(  s=10\right)&=&\frac{1}{d+2},\\
\Pr\left(  s=11\right)&=&\frac{d-1}{d+2}.
\end{eqnarray}
Moreover,
\begin{equation}
\Pr\left(   U=U_{i}\mid s=00\right)=
\left\{\begin{array}{lcr}
1& &i=d+2\\
0& &i\neq d+2
\end{array}
\right. ,
\end{equation}
\begin{equation}
\Pr\left(   U=U_{i}\mid s=01\right)=
\left\{\begin{array}{lcr}
1& &i=\overline{k}\\
0& &i\neq \overline{k}
\end{array}
\right. ,
\end{equation}
\begin{equation}
\Pr\left(   U=U_{i}\mid s=10\right)=
\left\{\begin{array}{lcr}
1& &i=\overline{k'}\\
0& &i\neq \overline{k}
\end{array}
\right. ,
\end{equation}
\begin{equation}
\Pr\left(   U=U_{i}\mid s=11\right)=
\left\{\begin{array}{lcr}
0& &i=\overline{k},\overline{k'}, d+2\\
\frac{1}{d-1}& &i\neq \overline{k},\overline{k'}, d+2
\end{array}
\right. ,
\end{equation}
then Bob's a posteriori uncertainty becomes
\begin{eqnarray}\label{Hpost2}
H_{post}\left(  U\right)  =\frac{d-1}{d+2}\log\left( d-1\right).
\end{eqnarray}
Equation (\ref{Hpost2}) can straightforwardly be generalized to strings of  $n+1$ qudits 
(with $n+3$ possible $s$ values) as
\begin{eqnarray}
H_{post}\left(  U\right)  =\frac{d-n}{d+2}\log\left(d-n\right).
\end{eqnarray}
It is easy to see that the uncertainty becomes $0$, when $d-n=1$, or when $n=d-1$.
Hence in order for Bob to have complete information on the unitary
transformation used, the number of qudits sent must be at least
equal to $d$.
\end{proof}

It is worth noting that the problem of encoding is
consequently solved as well.


\section{Efficiency of the Protocol}

Despite the fact that Alice and Bob may communicate reliably as
described above, the protocol is far from being efficient. Let us
consider the definition of efficiency for blind encoding in the
framework of communication. This definition closely follows the work
of Ref.\cite{cabello1}. It is well known that given a quantum system
the maximally attainable classical information $I$ (in independent
measurements) is bounded by the Holevo bound \cite{Keyl}
\begin{eqnarray}
I\leq S(\rho)-\sum_i p_{i}S(\rho_{i}).
\end{eqnarray}
Since each qudit sent and received by Bob may perfectly be distinguished by Bob
(as the states represent pure orthogonal states), he saturates the Holevo bound thus achieving,
with $d$ qudits, the maximal information 
$d\log_{2}d$ bits.
However, given a protocol, the actual amount of information that may
be shared between Alice and Bob may differ from this maximal value.
If we consider the shared information between Alice and Bob in the
previous section, an ideal channel would have allowed its
classical capacity \cite{Keyl} to be
\begin{eqnarray}
C:=\max I\left(  A:B\right)  =-\log\left(  \frac{1}{d+2}\right).
\end{eqnarray}
Defining efficiency to be the ratio of the perfect
channel capacity to the
maximal information of $d$ qudits, we get
\begin{eqnarray}
\frac{\log_{2}\left(  d+2\right)  }{d\log_{2}d}\leq1,
\end{eqnarray}
with equality only in the case for qubits ($d=2$). The ratio
decreases by increasing $d$. In order to achieve unit
efficiency we must force the numerator to be $d\log_{2}d$ or we must allow
Alice to use $d^{d}$ unitary operations. Intuitively this
may be understood as follows: since $d$ qudits are sent back and forth
between Bob and Alice, the exhaustive number of codewords that may
be shared between them would be $d^{d}$. Therefore, Alice must be
able to execute $d^{d}$  unitary  operations for encoding while Bob
must perfectly decode them. This means that Alice must construct
unique operations that the composite system of $d$ qudits (of differing
MUB) could actually discern.

We consider the composed system state of $d$ qudits
\begin{eqnarray}\label{Psi}
\left\vert \Psi\right\rangle \equiv\left\vert \psi_{t_1}^{1}\psi_{t_2}^{2}\ldots\psi_{t_d}^{d}\right\rangle
\end{eqnarray}
where the $\psi^i$ belongs to $d$ different MUB.
Vectors like (\ref{Psi}) span the space
of $d^{d}$ independent vectors. Then the encoding problem can be recast into the following form:
can we find an encoding operation on $\left\vert
\Psi\right\rangle $ such that $\left\vert
\psi_{t_l}^{l}\right\rangle $ gets shifted to $\left\vert
\psi_{t_l\oplus a}^{l}\right\rangle $ for any integers
$t_l\in[1,d]$ and $a\in[0,d-1]$? The answer to this question is
positive.

Let us first define the operation
$U_{\overline{d+1}}$ as the operation that does not shift the state
of the $(d+1)$th basis (which is not part of the composite system
$\left\vert \Psi\right\rangle $) but would shift the states of all
other qudits by $1$. If we wish to encode a value $a$ on a
particular qudit $i$, we need to ensure that while qudit $i$ gets
shifted by $a$, the others would not be affected by the shift. Hence we
could operate on the system $\left\vert \Psi\right\rangle$ the
operations $U_{\overline{i}}^{d-a}U_{\overline{d+1}}^{a}$ which
shifts the qudit $i$ by $a$ and the others by
$\left(  d-a+a\right) \operatorname{mod}d=0$. It is straightforward
to see that we could execute a similar recipe to any other qudits in
$\left\vert \Psi \right\rangle$. Hence, a general
encoding may be written as
\begin{eqnarray}\label{setU}
\left(  U_{\overline{1}}^{d-a}U_{\overline{d+1}}^{a}\right)  \left(
U_{\overline{2}}^{d-b}U_{\overline{d+1}}^{b}\right)\ldots\left(
U_{\overline
{d}}^{d-k}U_{\overline{d+1}}^{k}\right)
\left\vert \psi_{t_1}^{1}\psi_{t_2}^{2}\ldots\psi_{t_d}^{d}\right\rangle,
\end{eqnarray}
where $a,b,\ldots,k$ are
different values to encode onto the relevant qudits. Simple
observation tells us that we may encode in such a way all the $d^{d}$ codewords.

The remaining question is whether such a combination is unique; i.e.
despite the possible combinations for a string (subject to the
condition that no two qudits in the string share a basis), one set
of operations results in only one unique evolution of a string of states.
This is vital as a set of operations must be recognizable by such strings of qudits.
Let us write Eq.(\ref{setU}) as $A\left\vert \Psi\right\rangle \rightarrow\left\vert
\Phi\right\rangle$
and suppose that another operation $B$ gives
$B\left\vert \Psi\right\rangle \rightarrow\left\vert
\Phi\right\rangle$
then it obviously follows $A=B$. It goes without saying
that other strings do not yield the same
result as that of $\left\vert \Psi\right\rangle$. As long as Alice
and Bob agrees to codewords to be designated to each set of
operations, they may then have a faithful encoding/decoding procedure
which is of unit efficiency. This is important since the above
construction of encoding operations is done in
states of known bases, while Alice needs to use these operations on
unknown strings of qudits.


\subsection{The Qubit Case}

Consider Bob preparing two qubits states $\left\vert \psi_{t_1}^1\otimes\psi
_{t_2}^2\right\rangle $ with  $\left\vert \psi_{t_1}^1\right\rangle$, $\left\vert \psi_{t_2}^2\right\rangle$ eigenstates of Pauli operator $X$ and $Y\equiv XZ$ respectively. In constructing an operation
which may flip the states of both qubits, Alice makes use of Eq.(\ref{setU})
\begin{eqnarray}
\left(  U_{\overline{x}}^{2-a}U_{\overline{z}}^{a}U_{\overline{y}}^{2-b}
U_{\overline{z}}^{b}\right)  ^{\otimes 2}\left\vert
\psi_{t_1}^1\otimes\psi _{t_2}^2\right\rangle \rightarrow\left\vert
\psi_{t_1\oplus a}^1 \otimes  \psi_{t_2\oplus b}^2
\right\rangle,
\end{eqnarray}
with $a,b\in[0,1]$.

We can readily convince ourselves of the above by noticing
\begin{eqnarray}
\left(  U_{\overline{x}}^{2-a}U_{\overline{z}}^{a}\right)   &  \equiv X^{2-a}Z^a,\\
\left(  U_{\overline{y}}^{2-b}U_{\overline{z}}^{b}\right)   &  \equiv Y^{2-b}Z^b.
\end{eqnarray}

\subsection{The Qutrit Case }

Consider Bob preparing three qutrits states $\left\vert
\psi_{1}\otimes \psi_{2}\otimes\psi_{3}\right\rangle $ with the
integer indices denoting different MUB. In constructing operations
which may flip the states of qutrits, Alice makes use of Eq.(\ref{setU})
\begin{align}
& \left[\left( U_{\overline{1}}^{3-a}U_{\overline{4}}^{a}\right)
\left(
U_{\overline{2}}^{3-b}U_{\overline{4}}^{b}\right)  \left(  U_{\overline{3}
}^{3-c}U_{\overline{4}}^{c}\right) \right]^{\otimes 3}
\left\vert \psi_{t_1}^1\otimes\psi_{t_2}^2
\otimes\psi_{t_3}^3\right\rangle \cr &
\rightarrow\left\vert
\psi_{t_1\oplus a}^1\right\rangle \left\vert \psi_{t_2\oplus
b}^2\right\rangle \left\vert \psi_{t_3\oplus c}^3\right\rangle,
\end{align}
with $a,b,c\in[0,1,2]$.

The unitary operations $U_{\overline{i}}$ corresponds to the well
known shift/error operators for qutrits \cite{cerf}.
A table for the explicit evolution of the various possible states
Bob may prepare under Alice's transformation is referred to in
\cite{qutrit}.


\section{Conclusion}

We have considered the problem of blind encoding classical information into quantum
states belonging to a maximal set of MUB for 
systems whose dimensions equal to a prime number $d$. We noted
that trivial encoding is essentially forbidden due to the inability
of reliably shifting an unknown arbitrary qudit by unitary operations.
We proved this in a no-go theorem which is a generalisation
of specific case treated in Ref.\cite{qutrit}. On the other hand, Bob
cannot reliably decode the information content of a qudit unless he
actually prepares and then measures a string of $d$ qudits.
We provided a simple information theoretic proof
for this lemma.

We then noticed that while $d$ is the minimum number of qudits that
must be sent to Alice for reliable decoding, the available
generalised Pauli operations do not provide an
efficient encoding protocol. Unit efficiency may
be achieved with the protocol we proposed for the relevant encoding
and decoding procedure. Two examples of application of this protocol has been explicitly shown.

The developed approach paves the way for further studies of two way quantum communication with channels of generic dimension and may be useful for cryptographic tasks.


\section*{Acknowledgements}

J.S.S. is grateful to the Faculty of Science of
IIUM for the facilities provided to him in undertaking his
doctorate programme.
S.M. acknowledges financial support from European Union through the integrated project ``QAP" 
(IST-FET FP6-015848).


\end{document}